\documentclass[lettersize,journal]{IEEEtran}

\usepackage{cite}
\usepackage{amsmath,amssymb,amsfonts}
\usepackage{graphicx}
\usepackage{textcomp}
\usepackage{xcolor}
\usepackage{authblk}

\usepackage{booktabs}
\usepackage{epsfig}
\usepackage{latexsym}
\usepackage{multirow}
\usepackage{stfloats}
\usepackage{epstopdf}
\usepackage{color}  
\usepackage{tabularx} 
\usepackage{enumerate}
\usepackage{array}
\graphicspath{{./Figures/}}
\usepackage{color}
\usepackage{bbm}
\usepackage{caption}
\usepackage{bm}
\usepackage[tight,footnotesize]{subfigure}
\usepackage{balance}
\usepackage{mathrsfs}
\usepackage{verbatim}
\allowdisplaybreaks[4]
\usepackage{dsfont}
\usepackage{verbatim}
\usepackage{tikz}
\usepackage{setspace}
\usepackage{diagbox}
\usepackage[framemethod=tikz]{mdframed}
\usepackage{multicol}
\usepackage{environ}
\usepackage{tikz}
\usepackage{stfloats}
\usepackage{algpseudocode}
\usepackage{graphics}
\usepackage{epsfig}
\usepackage{amsthm}
\usepackage{authblk}
\usepackage{enumitem}
\usepackage{makecell}
\usepackage{courier}

\ifCLASSOPTIONcompsoc
\usepackage[caption=false, font=normalsize, labelfont=sf, textfont=sf]{subfig}
\else
\usepackage[caption=false, font=footnotesize]{subfig}

\usepackage{bm}
\def\BibTeX{{\rm B\kern-.05em{\sc i\kern-.025em b}\kern-.08em
    T\kern-.1667em\lower.7ex\hbox{E}\kern-.125emX}}
\begin{document}

\title{A Geography-Inspired and Self-Adaptive Clustering Algorithm: A Study in Channel Measurement}

\author{Yiqin~Wang,
and Chong~Han,~\IEEEmembership{Senior~Member,~IEEE}%
\thanks{Yiqin Wang is with Terahertz Wireless Communications (TWC) Laboratory, Shanghai Jiao Tong University, China (Email: wangyiqin@sjtu.edu.cn).}%
\thanks{Chong Han is with the Terahertz Wireless Communications (TWC) Laboratory and also the Cooperative Medianet Innovation Center (CMIC), School of Information Science and Electronic Engineering, Shanghai Jiao Tong University, China (Email: chong.han@sjtu.edu.cn).}%
}

\maketitle
\begin{abstract}
The phenomenon that multi-path components (MPCs) arrive in clusters has been verified by channel measurements, and is widely adopted by cluster-based channel models. As a crucial intermediate processing step, MPC clustering bridges raw data in channel measurement and cluster characteristics for channel modeling. In this paper, a physical-interpretable and self-adaptive MPC clustering algorithm is proposed, which can locate both single-point and wide-spread scatterers without prior knowledge. Inspired by the concept in geography, a novel metaphor that interprets features of MPC attributes in the power-delay-angle profile (PDAP) as topographic concepts is developed. In light of the interpretation, the proposed algorithm disassembles the PDAP by constructing contour lines and identifying characteristic points that indicate the skeleton of MPC clusters, which are fitted by analytical models that associate MPCs with physical scatterer locations. Besides, a new clustering performance index, the power gradient consistency index, is proposed. Calculated as the weighted Spearman correlation coefficient between the power and the distance to the center, the index captures the intrinsic property of MPC clusters that the dominant high-power path is surrounded by lower-power paths.
The performance of the proposed algorithm is analyzed and compared with the counterparts of conventional clustering algorithms based on the channel measurement conducted in an outdoor scenario. The proposed algorithm performs better in average Silhouette index and weighted Spearman correlation coefficient, and the average root mean square error (RMSE) of the estimated scatterer location is 0.1~m.
\end{abstract}

\begin{IEEEkeywords}
Clustering algorithm, Localization, Channel measurement, Channel modeling.
\end{IEEEkeywords} 
\section{Introduction}

\par Since the phenomenon that multi-path components (MPCs) arrive in clusters is verified by channel measurements, cluster-based channel models have been widely adopted~\cite{saleh1987statistical,molisch2002virtual,kunisch2003ultra,molisch2006cost259,liu2012cost,Kyosti2008WINNER,3gpp36873,3gpp38901}. As a result, clusters have been regarded as an important feature in cluster-based channel modeling, where cluster-related parameters such as cluster number, position, and inter- and intra-cluster spreads are characterized. In the study of wireless channels, MPC clustering serves as a crucial intermediate processing step that bridges raw measurement data and channel modeling characteristics.

\begin{figure*}[!tb]
    \centering
    \includegraphics[width=\linewidth]{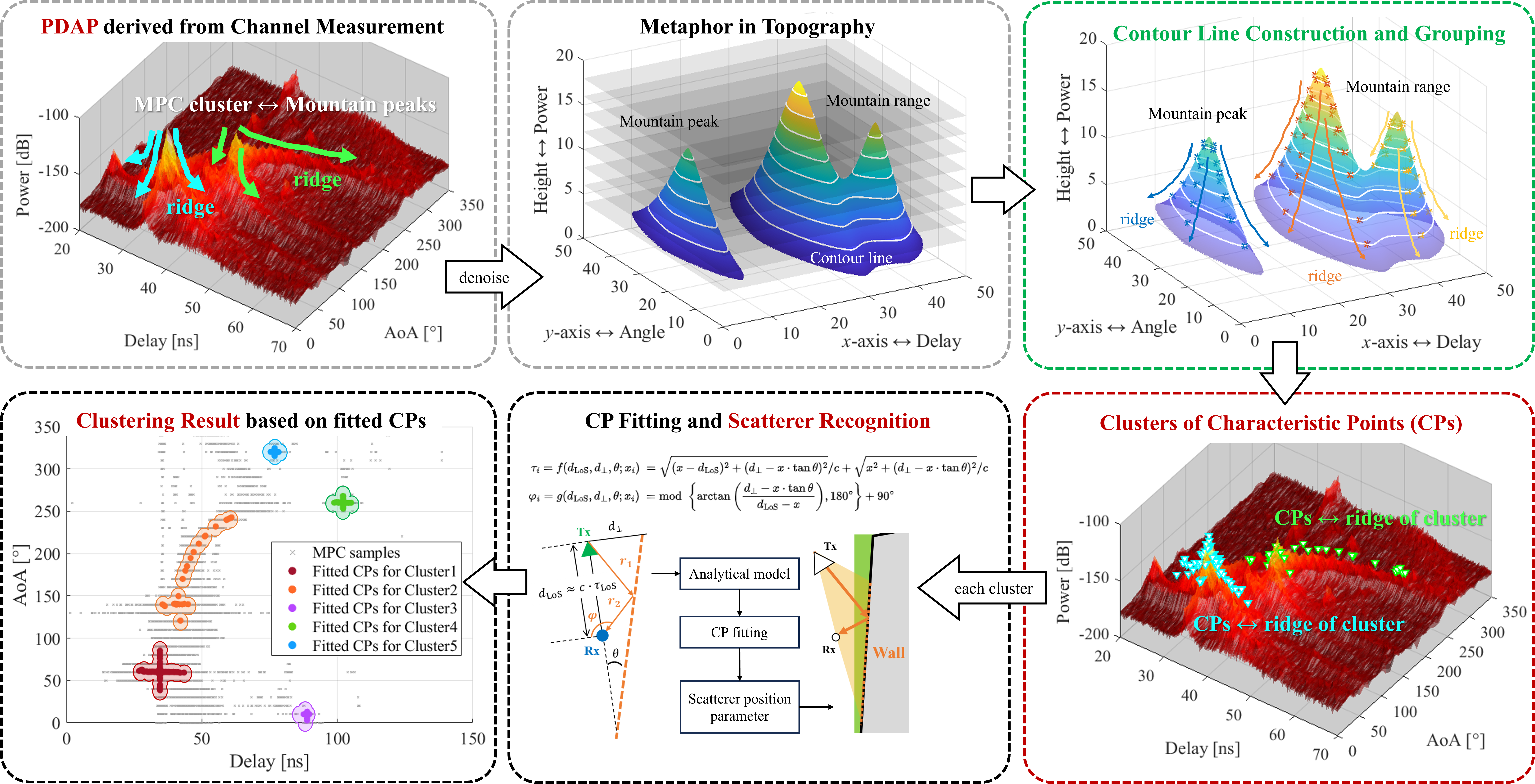}
    \caption{Schematic diagram of the proposed clustering algorithm.}
    \label{fig:flow_diagram}
\end{figure*}

\par There are three existing definitions of cluster in channel modeling~\cite{sun2021study}. Among them, one defines clusters as groups of MPCs with similar temporal evolution between different snapshots of the channel. This concept is specially defined for time-varying channels and gives rise to MPC cluster tracking algorithms~\cite{wang2017framework,huang2020trajectory,sun2021study}. Instead, in the scope of this work, we focus on clusters in one snapshot of a channel, which is commonly adopted in the field of MPC clustering~\cite{chen2021framework,he2017kpd,he2017automatic,hu2020novel,zhou2022machine,qiao2020novel,huang2018novel,huang2019power,hanpinitsak2014clustering}.

\par On one hand, a cluster can be defined as a group of MPCs with similar parameters such as delay and angles. In light of the definition, various multi-path component distances (MCDs)~\cite{chen2021channel} in terms of MPC attributes (delay, angle, and power) are developed to accommodate existing clustering algorithms like k-means~\cite{macqueen1967some} and density-based spatial clustering of applications with noise (DBSCAN)~\cite{ester1996dbscan}. For instance, a machine learning (ML)-based MCD, under the framework of Mahalanobis-distance metric, is proposed in~\cite{chen2021framework}. A power-weighted MCD is proposed in Kernel power-density (KPD) based clustering algorithm, which combines the ideas of k-means and DBSCAN~\cite{he2017kpd,he2017automatic}. Besides, ML methods are adapted for MPC clustering based on spectral clustering~\cite{hu2020novel} or Gaussian mixture model (GMM) clustering~\cite{zhou2022machine,qiao2020novel}. Furthermore, by transforming the delay-angular profile into an image, MPC clustering can also leverage image processing operations such as erosion and dilation~\cite{huang2018novel,huang2019power}.

\par On the other hand, an alternative definition regards a cluster as a group of MPCs arising from a cluster of physical objects, which associates MPCs with physical objects (or scatterers). The definition is widely used in geometry-based stochastic channel models (GBSMs). As a result, corresponding clustering algorithms require the knowledge of physical objects (or scattering points) obtained by deterministic methods like ray-tracing~\cite{hanpinitsak2014clustering}.

\par While MPC clusters are expected to inherently satisfy both definitions simultaneously, the field still lacks an MPC clustering algorithm that can group MPCs with similar parameters and concurrently associate them with physical scatterers without prior knowledge.
To fill this research gap, in this paper, a geography-inspired, physical-interpretable, and self-adaptive MPC clustering algorithm is proposed. The main contributions of this work are summarized as follows.

\begin{itemize}
    \item Inspired by the concept in geography, we propose a novel metaphor that interprets features of MPC attributes in the power-delay-angle profile (PDAP) as topographic concepts such as mountain peaks and mountain ranges.
    In light of the interpretation, the proposed algorithm disassembles the PDAP by constructing contour lines with all MPC attributes, i.e., delay, angle, and power. Contour lines, with points in delay and angles, are hierarchically organized in a tree structure, whose layers are mapped to power values. As a result, the number of clusters can be determined by back-tracing the contour line tree.
    \item The proposed algorithm surpasses conventional clustering algorithms in channel data processing by autonomously localizing both single-point and wide-spread scatterers without prior knowledge, and thus embedding physical significance into the clustering result. To be specific, the algorithm first identifies characteristic points (CPs) on contour lines, which indicate the ``ridge'' of MPC clusters, analogous to the ridge of mountains in topography. CPs, in terms of delay and angle, direct the clustering of MPCs and are fitted by analytical models that associate MPCs with physical scatterer locations.
    \item We propose a new clustering performance index, named as the power gradient consistency index, which evaluates the intrinsic property of MPC clusters that the dominant high-power path is surrounded by lower-power paths. The index assesses whether the power distribution in each cluster adheres to a radially decreasing trend from the center to the periphery, by evaluating the weighted Spearman correlation coefficient between the power and the distance to the center.    
    \item Based on the channel measurement conducted in an outdoor street, the performance of the proposed algorithm is comprehensively compared with the counterparts of conventional algorithms. Besides the decimeter-level localization precision and the better performance in terms of average Silhouette index and weighted Spearman correlation coefficient, the proposed algorithm is also free of user-specific parameters, and is insensitive to the valid MPC power threshold in post-processing of channel measurement data. Furthermore, the clustering algorithm is valid for general frequencies and scenarios, and is not limited to the application in channel measurement.
\end{itemize}

\par The remainder of this paper is organized as follows. In Section~\ref{sec:algorithm}, the proposed algorithm is described in detail. In Section~\ref{sec:campaign_and_result}, we conduct channel measurements in an outdoor scenario at 300~GHz and verify the performance, both in MPC clustering and scatterer identification, of the proposed algorithm based on the measurement data. In Section~\ref{sec:compare}, The clustering performance of the proposed algorithm is compared with the counterparts of conventional clustering algorithms, in terms of the Silhouette index and the proposed power gradient consistency index. Finally, the paper is concluded in Section~\ref{sec:conclusion}.

\section{The Proposed Algorithm} \label{sec:algorithm}

\subsection{Overview}
\par The flow diagram of the proposed clustering algorithm is shown in Fig.~\ref{fig:flow_diagram}.
Based on the PDAP obtained from the channel measurement, the first step is denoising. Due to the fluctuation of noise power, for simplicity, power thresholding is adopted to filter out samples below the maximum noise level.

\par Based on the intrinsic property of MPC clusters, we propose a geographical metaphor that maps the features of MPC attributes in the PADP to concepts in topography.
To be specific, delay and angle are used to indicate the position of MPC samples. The power of each sample denotes its height. Therefore, one MPC cluster is analogous to a mountain peak, while the superposition of contiguous clusters forms a mountain range.

\par The key idea of the proposed algorithm is to characterize the skeleton of MPC clusters as the ridge of the mountain peaks, which is utilized to guide the MPC clustering and recognize physical scatterers. Samples on the cluster ridge are defined as characteristic points, which are derived as convex points on contour lines, and are divided into clusters based on the hierarchical organization of contour lines in a tree structure.

\begin{figure}[!t]
    \centering
    \includegraphics[width=\linewidth]{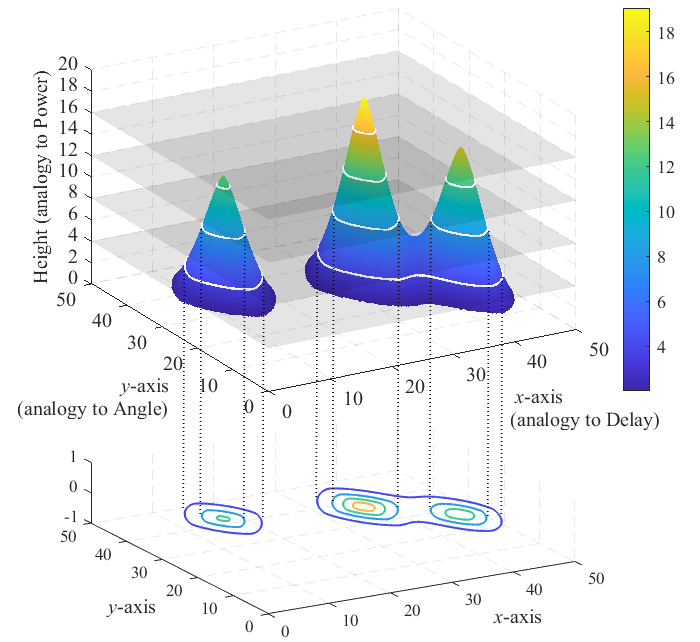}
    \caption{The example of contour lines.}
    \label{fig:contour_line_example}
\end{figure}

\begin{figure}[!t]
    \centering
    \includegraphics[width=0.85\linewidth]{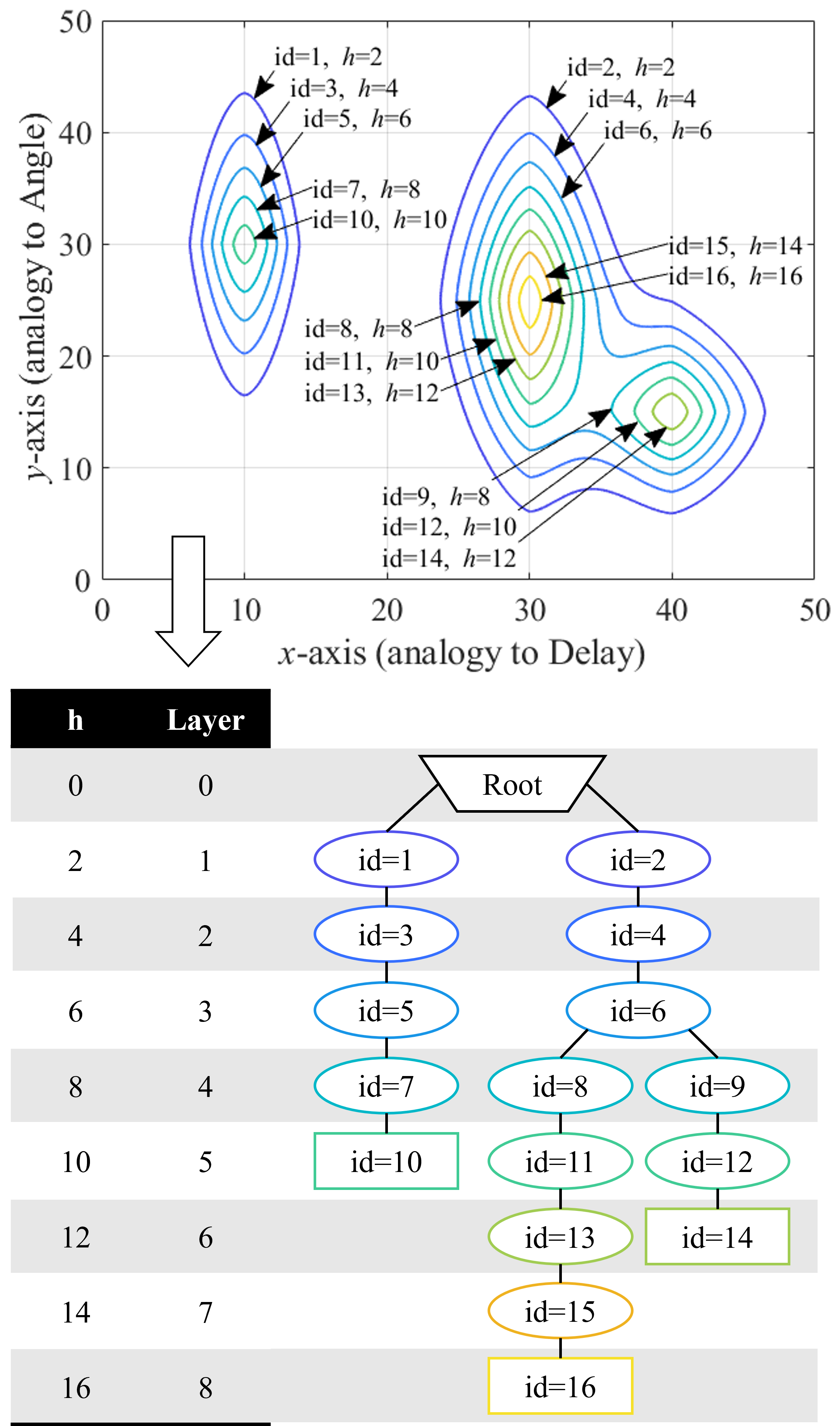}
    \caption{The example of a contour line tree.}
    \label{fig:contour_line_tree_example_2}
\end{figure}

\begin{figure}[!t]
    \centering
    \includegraphics[width=\linewidth]{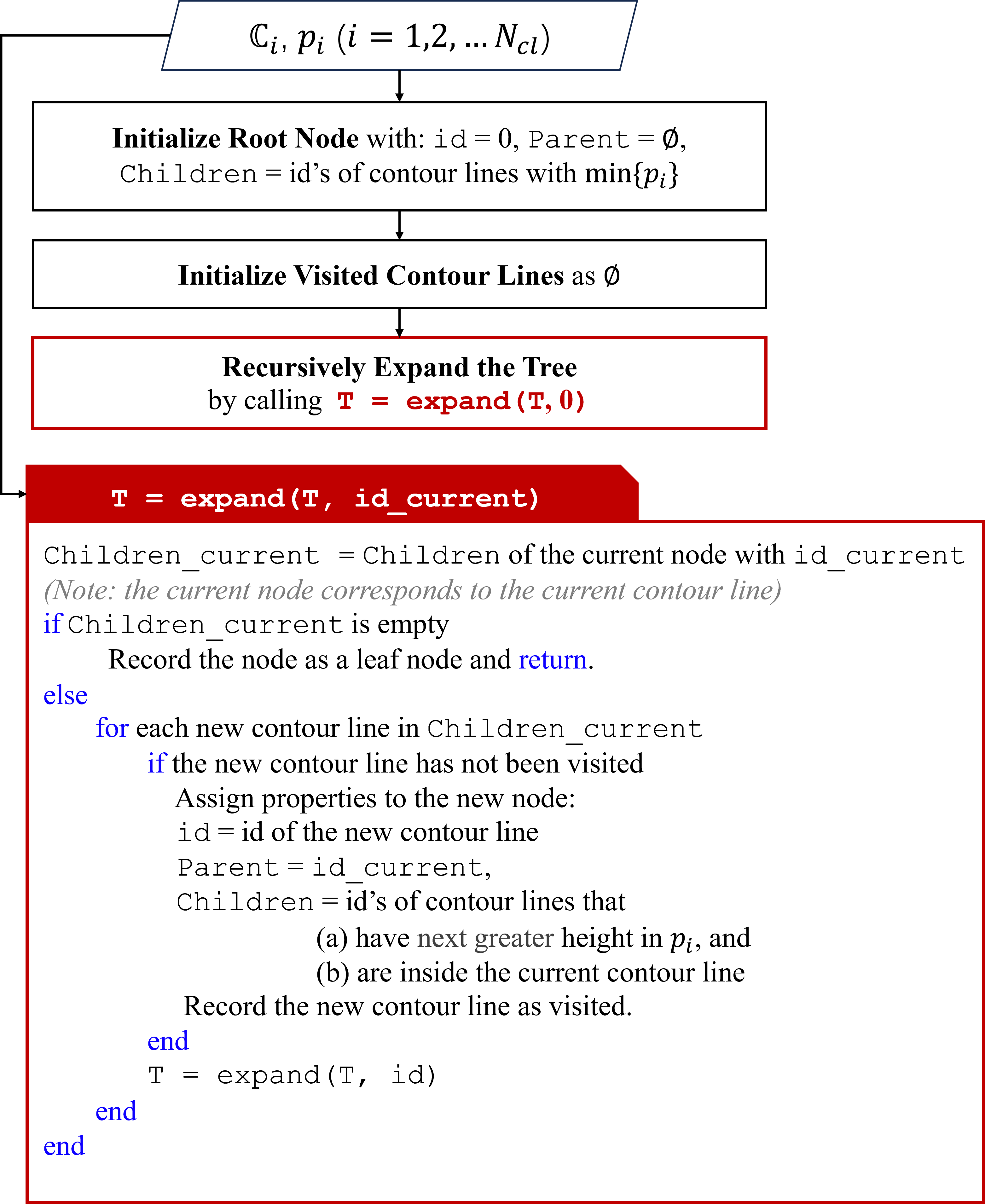}
    \caption{The algorithm of contour line tree construction.}
    \label{fig:algorithm_tree_construction}
\end{figure}

\subsection{Contour Line Construction and Grouping}
\par Fig.~\ref{fig:contour_line_example} shows an example of contour lines in $x$ and $y$ axes by sweeping the value of height. The numerical values in the figure are for demonstration purpose and have no substantive meaning. The same method is applied to derive contour lines of PDAP. Fig.~\ref{fig:contour_line_example} illustrates the example of an independent cluster and a group of two contiguous clusters. Each contour line is bounded with one power value.

\par We denote the number of contour lines as $N_{cl}$, the $i$-th contour line as $\mathbb{C}_i=\{(\tau,\varphi)\}$ which contains samples with delay $\tau$ and angle $\varphi$, and the power of the $i$-th contour line as $p_i$. After deriving contour lines $\mathbb{C}_i$ with corresponding $p_i$ ($i=1,2,...,N_{cl}$), we employ the following algorithms to assign these contour lines into groups and determine the number of clusters.

\par To start with, we transfer contour lines into a tree, the data structure that features the hierarchical organization and efficient operations like traversal and back-tracing. Each contour line is represented by a unique id.
The layers of the tree correspond to the order of sweeping heights, with the root at Layer~0.
An example of contour lines and the result contour line tree is shown in Fig.~\ref{fig:contour_line_tree_example_2}.

\par In the contour line tree, each node corresponds to a contour line in $\mathbb{C}_i$. Each node has three properties, i.e., the value of the node (\texttt{id}), the value of the parent node (\texttt{Parent}), and values of children (\texttt{Children}). The \texttt{Children} of a contour line satisfy the following conditions. First, the height of the child contour line is the next greater value in $p_i$. Second, the child contour line lies inside the current contour line, which can be determined by $\mathbb{C}_i=\{(\tau,\varphi)\}$.
The tree is constructed by recursively expanding the node with contour lines traversed in ascending order of height. Details of the algorithm are summarized in Fig.~\ref{fig:algorithm_tree_construction}. 

\par Furthermore, we assign contour lines into groups and determine the number of clusters. Given the contour line tree, we can efficiently obtain all root-to-leaf paths by back-tracing the contour line tree from each leaf node to the root node. As a result, the nodes, i.e., contour lines, on each obtained path form a group. For instance, in Fig.~\ref{fig:contour_line_tree_example_2}, three groups, with id's (1,3,5,7,10), (2,4,6,8,11,13,15,16) and (2,4,6,9,12,14) are found. By this step, we are able to determine the number of clusters, which is equal to the number of contour line groups, since the leaf node indicates the contour line with the largest power, which refers to the peak of a cluster.

\subsection{Characteristic Point Recognition}
\par In this part, we identify CPs on each contour line.
Since contour lines are closed curves, we define CPs as convex points on contour lines. Before that, two steps are required. First, due to the fluctuation of experiment data, the contour line is smoothed by interpolation. Second, the coordinates on the delay axis and the angle axis are normalized based on the range of delays and angles on the contour line.
\par We denote the interpolated contour line with $N_{\rm smooth}$ ordered points as $\mathbb{C_{\rm smooth}}=\{P_{m}(\tau_{m}',\varphi_{m}')\}_{m=1}^{N_{\rm smooth}}$ ($P_{1}=P_{N_{\rm smooth}+1}$). $\tau_{m}'$ and $\varphi_{m}'$ are normalized delays and angles of the $m$-th point. Then, we traverse the contour line by examining the previous point $P_{m-1}$, the current point $P_{m}$, and the next point $P_{m+1}$ at a time. The current point is regarded as a CP if (a) the angle $\angle P_{m-1}P_{m}P_{m+1}$ is a local minimum for $m=1,2,...,N_{\rm smooth}$, and (b) the current point $P_{m}$ is a convex point, i.e.,
\begin{equation}
    \overrightarrow{P_{m}-P_{m-1}} \times \overrightarrow{P_{m+1}-P_{m}} > 0,
\end{equation}
where $\times$ denotes the cross product between vectors.

\par
\begin{figure}[!t]
    \centering
    \includegraphics[width=\linewidth]{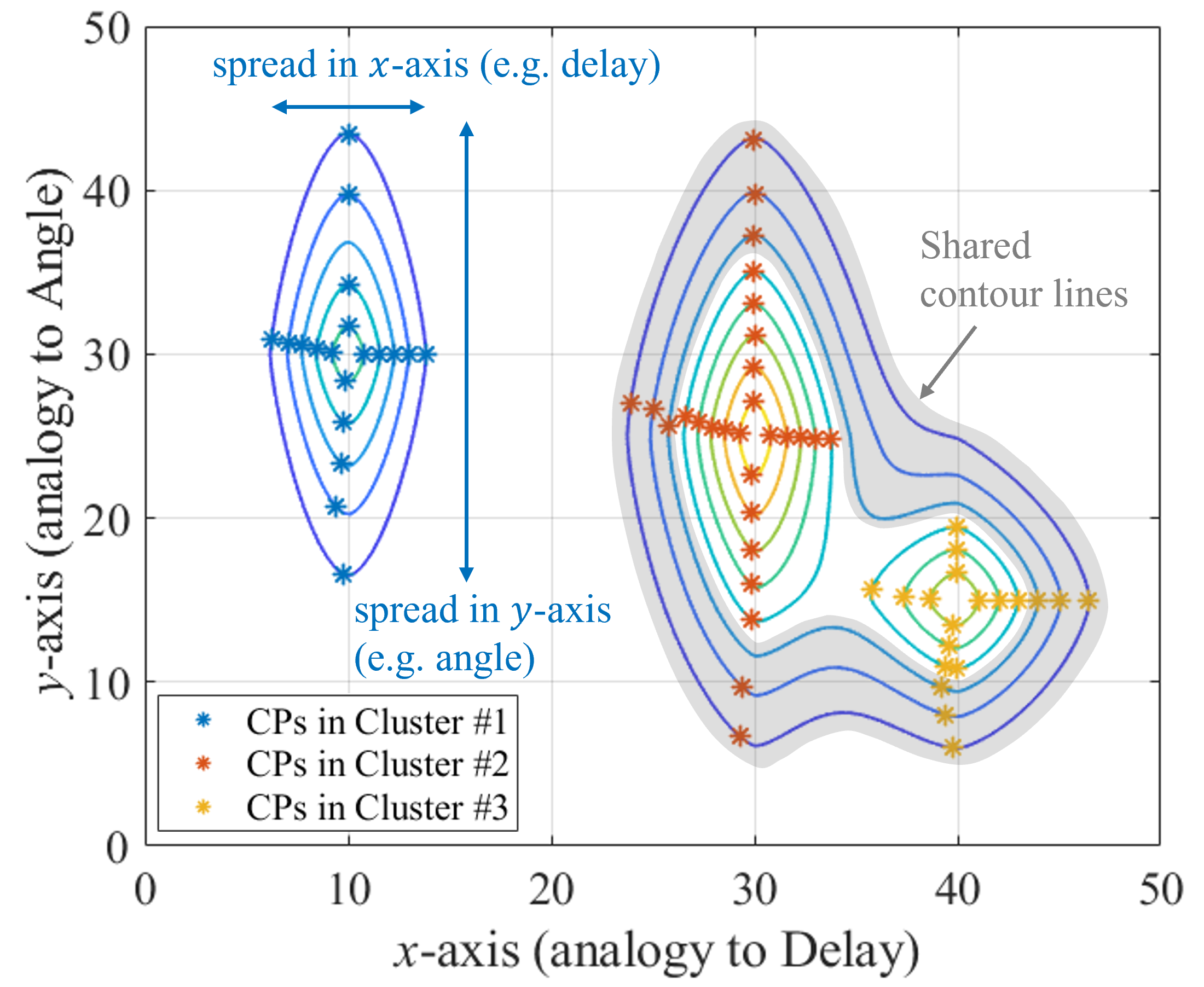}
    \caption{The example of characteristic points.}
    \label{fig:characteristic_point_example}
\end{figure}

 Analogous to the ridge of mountains in geomorphology, CPs are samples on the ``ridge'' of clusters in the PDAP.
In the simplest case, a cluster is a result from an object with one strong reflection point, and thus the cluster's ridge is equivalent to the directions of power spread along axes of delay and angle. In this case, ideally, CPs form two perpendicular straight lines intersected at the position of the dominant MPC, as the example in Fig.~\ref{fig:characteristic_point_example}.
Besides, in the experiment elaborated later in Section~\ref{sec:campaign_and_result}, a cluster is observed which is generated by continuous scattering points from a wide-spread object like a wall. In this case, the ``ridge'' of the cluster is curved in the delay-angle coordinates.

\par After deriving CPs for each contour line, we need to distribute CPs into clusters and, in the end, guide the scatterer recognition and MPC clustering. Fig.~\ref{fig:characteristic_point_example} also shows the result of CP clusters.
First of all, we need to identify, if exist, the greatest common contour lines between groups.
To be specific, if a contour line group does not share any contour line with other groups, CPs on contour lines in this group are assigned to the same cluster.
On the other hand, for contiguous clusters, it is likely to result in contour line groups that share a part of contour lines. For example, as shown in Fig.~\ref{fig:contour_line_tree_example_2}, group (2,4,6,8,11,13,15,16) and (2,4,6,9,12,14) share contour lines with id's 2, 4 and 6. In this case, CPs on unshared contour lines, (8,11,13,15,16) and (9,12,14) are first assigned to their own clusters, respectively. Then, CPs on shared contour lines are distributed to the nearest cluster. In this way, the ridges, characterized by CPs, of two contiguous clusters can be distinguished and remain independent.

\subsection{Clustering and Scatterer Recognition based on CPs}
\par CPs in the same cluster are utilized to indicate the cluster ridge, and thus guide the MPC clustering and recognize physical scatterers that result in the cluster. By this step, CPs have been extracted and clustered as the samples on the ridge of clusters. Therefore, the proposed algorithm assigns MPCs to the nearest CP cluster.

\par As analyzed before, for the MPC cluster resulting from an object with one strong reflection point, CPs of this cluster can be fitted by two perpendicular straight lines intersected at the position of the dominant MPC. One one hand, the physical scatterer of the cluster can be recognized by the position, i.e., delay and angle, of the intersected point.
One the other hand, one MPC joins a cluster if any CP in the cluster has the minimum distance to the MPC. Delays and angles are normalized to calculate the distance between CPs and MPCs.

\begin{figure}[!t]
    \centering
    \includegraphics[width=0.5\linewidth]{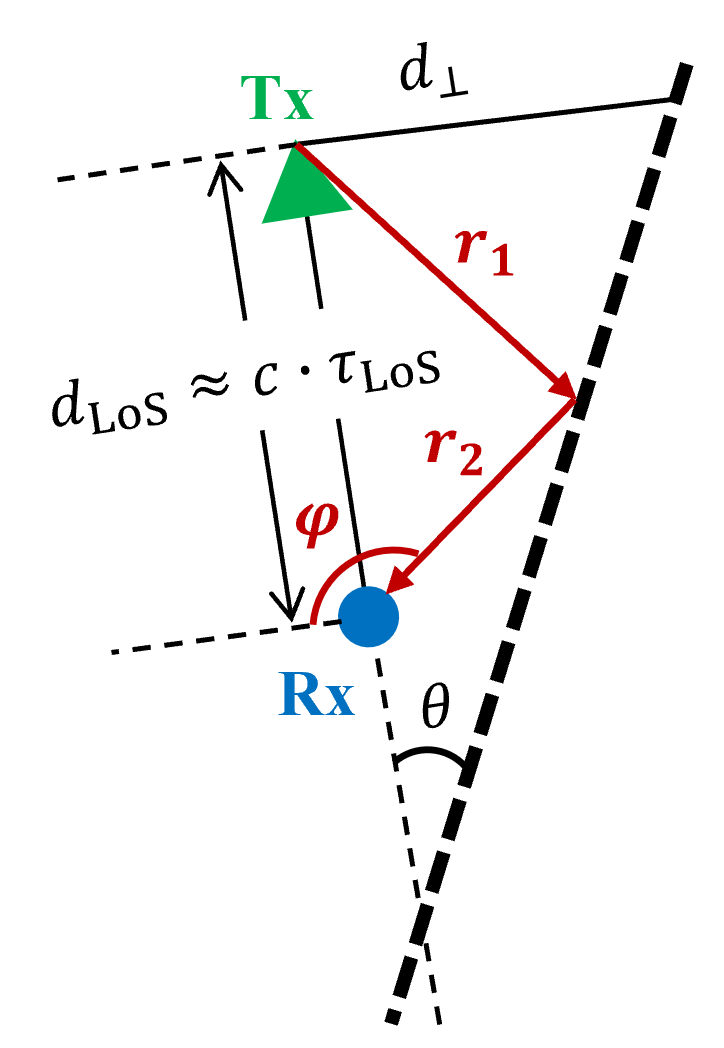}
    \caption{The analytical model of continuous scattering on a wide-spread object.}
    \label{fig:analytical_model}
\end{figure}


\par For the MPC cluster generated by continuous scattering points from a wide-spread object like a wall, we can also observe the continuous change of delay, AoA and power of scattering rays among PDAPs. Therefore, CPs follow a curved model in terms of delays and angles. CPs derived from the contour line group are fitted to the analytical model as shown in Fig.~\ref{fig:analytical_model} and described as
\begin{subequations}
\begin{align}
\begin{split}
    \tau_i =& f(d_{\rm LoS},d_{\perp},\theta;x_i)\\
    =& \sqrt{(x-d_{\rm LoS})^2+(d_{\perp} - x\cdot\tan\theta)^2}/c\\
    &+ \sqrt{x^2+(d_{\perp} - x\cdot\tan\theta)^2}/c,
\end{split}\\
\begin{split}
    \varphi_i =& g(d_{\rm LoS},d_{\perp},\theta;x_i)\\
    =&\mod\left\{\arctan\left(\frac{d_{\perp} - x\cdot\tan\theta}{d_{\rm LoS}-x}\right),180^\circ\right\}+90^\circ,
\end{split}
\end{align}
\end{subequations}
where the $i$-th CP in the cluster is represented by delay $\tau_i$ and AoA $\varphi_i$ ($i=1,2,...,N_{\rm CP}$). $c$ is the speed of light, $d_{\rm LoS}$ is the LoS distance between Tx and Rx, $d_{\perp}$ and $\theta$ determine the position of the scatterer as illustrated in Fig.~\ref{fig:analytical_model}. $x_i$ is the hidden variable that implicitly defines the relation between $\tau_i$ and $\varphi_i$.

\par Parameters ($d_{\rm LoS}$, $d_{\perp}$, $\theta$) and hidden variables $x_i$'s are estimated during the optimization. In view of the noise in the measurement data, the random sample consensus (RANSAC) paradigm~\cite{fischler1981RANSAC} is utilized to remove outliers in CPs. In particular, during each iteration of RANSAC, $N_s=10$ samples are randomly chosen to estimate parameters ($\hat{d}_{\rm LoS}$, $\hat{d}_{\perp}$, $\hat{\theta}$). Based on these parameters, we optimize $\hat{x}_i$ for each CP ($\tau_i$, $\varphi_i$) ($i=1,2,...,N_{\rm CP}$), where the error function for each CP sample is calculated as
\begin{equation}
\begin{split}
    e_i =& [ (\tau_i - f(\hat{d}_{\rm LoS},\hat{d}_{\perp},\hat{\theta};x_i))^2 \\
    &+ (\varphi_i - g(\hat{d}_{\rm LoS},\hat{d}_{\perp},\hat{\theta};x_i))^2 \\
    &+ w_{\rm prior}\cdot(\hat{d}_{\rm LoS} - d_{\rm LoS,prior})^2 ]^{\frac{1}{2}},
\end{split}
\end{equation}
where $d_{\rm LoS,prior}$ is the prior knowledge of estimated LoS distance, obtained from $c$ times the the arrival time of the strongest MPC in the PDAP. $w_{\rm prior}=2$ is the weight of the prior knowledge.
Then, if at least 30\% samples are regarded as inliers, i.e., has $e_i$ under the error threshold, with estimated parameters ($\hat{d}_{\rm LoS}$, $\hat{d}_{\perp}$, $\hat{\theta}$) and optimized $\hat{x}_i$ ($i=1,2,...,N_{\rm CP}$), we calculate the average error of inliers. The best inliers are updated if the average error is the minimum during the iterations.
After adequate iterations, the best inliers are used to estimate parameters ($d_{\rm LoS}$, $d_{\perp}$, $\theta$) and hidden variables $x_i$'s. Note that the number of iterations is determined to ensure that the choice of $N_s$ CP samples in these iterations covers the combination of $N_s$ out of $N_{\rm CP}$ CPs as much as possible. To decrease the number of iterations, the choice of $N_s$ samples is improved. To be specific, to improve the estimation performance, instead of the random selection, we adapt the partitioned sampling strategy, i.e., $N_s$ samples are selected across partitions of CPs. In this case, 10000 times of iteration is adequate.

\par Based on the estimated parameters ($d_{\rm LoS}$, $d_{\perp}$, $\theta$) and the analytical model illustrated in Fig.~\ref{fig:analytical_model}, continuous scattering points on a wide-spread object can be reconstructed. Therefore, the proposed algorithm can reconstruct not only a single-point scatterer, but also wide-spread, continuous scatterers like a wall, which provides significant physical evidence for the clustering result.

\section{Channel Measurement and Result} \label{sec:campaign_and_result}

\par In this section, we introduce the channel measurement in the outdoor street, and demonstrate the scatterer recognition and clustering result of the proposed algorithm based on the measurement data.

\begin{figure}[!t]
    \centering
    \includegraphics[width=0.85\linewidth]{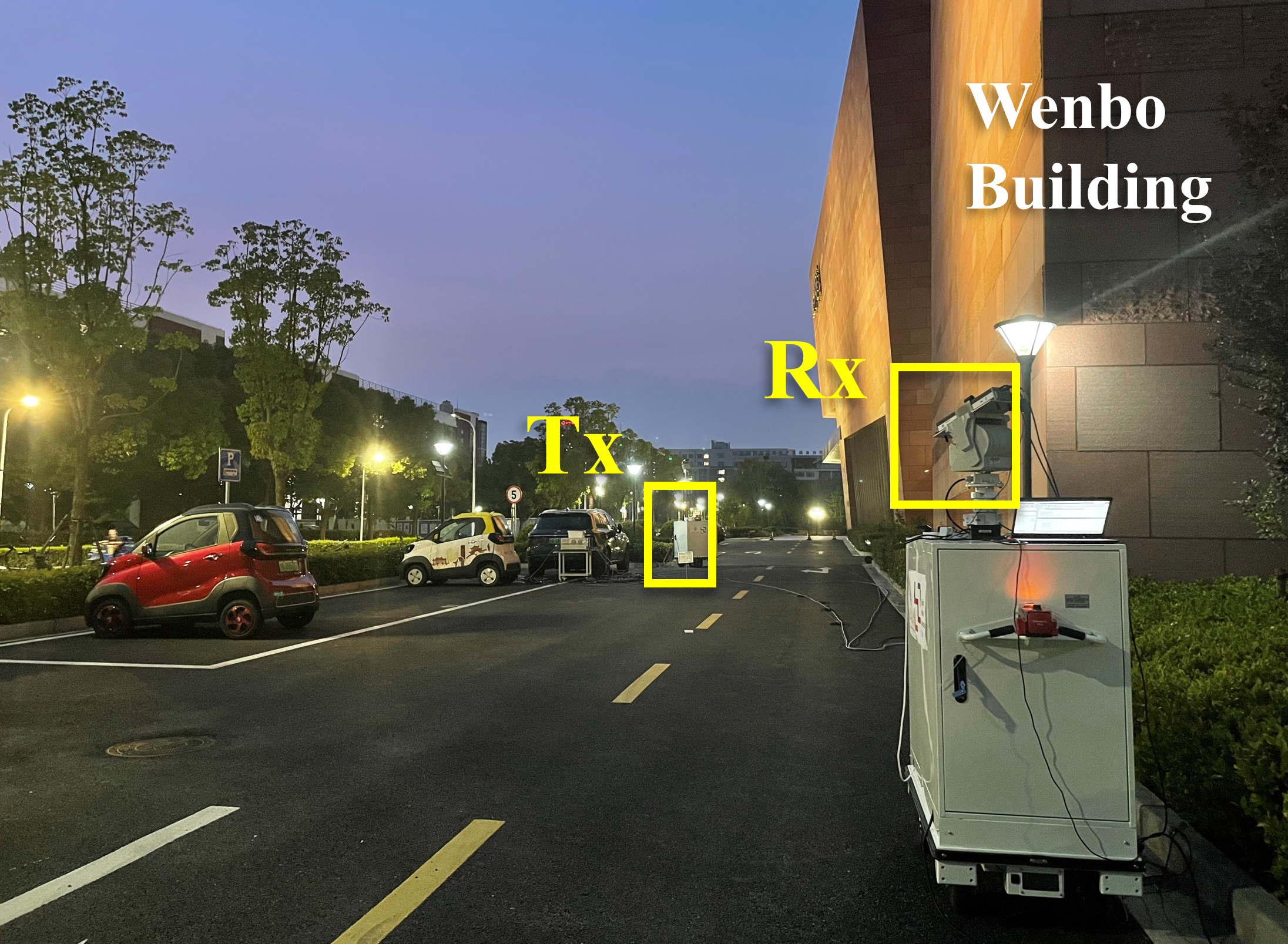}
    \caption{Photo of the outdoor street.}
    \label{fig:environment}
\end{figure}

\begin{figure}[!t]
    \centering
    \includegraphics[width=0.8\linewidth]{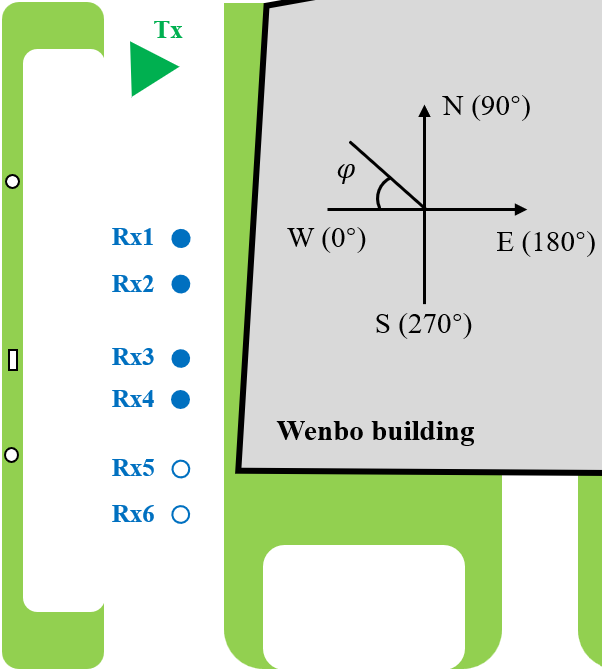}
    \caption{Measurement deployment in the outdoor street.}
    \label{fig:deployment}
\end{figure}

\begin{figure}[!t]
    \centering
    \includegraphics[width=\linewidth]{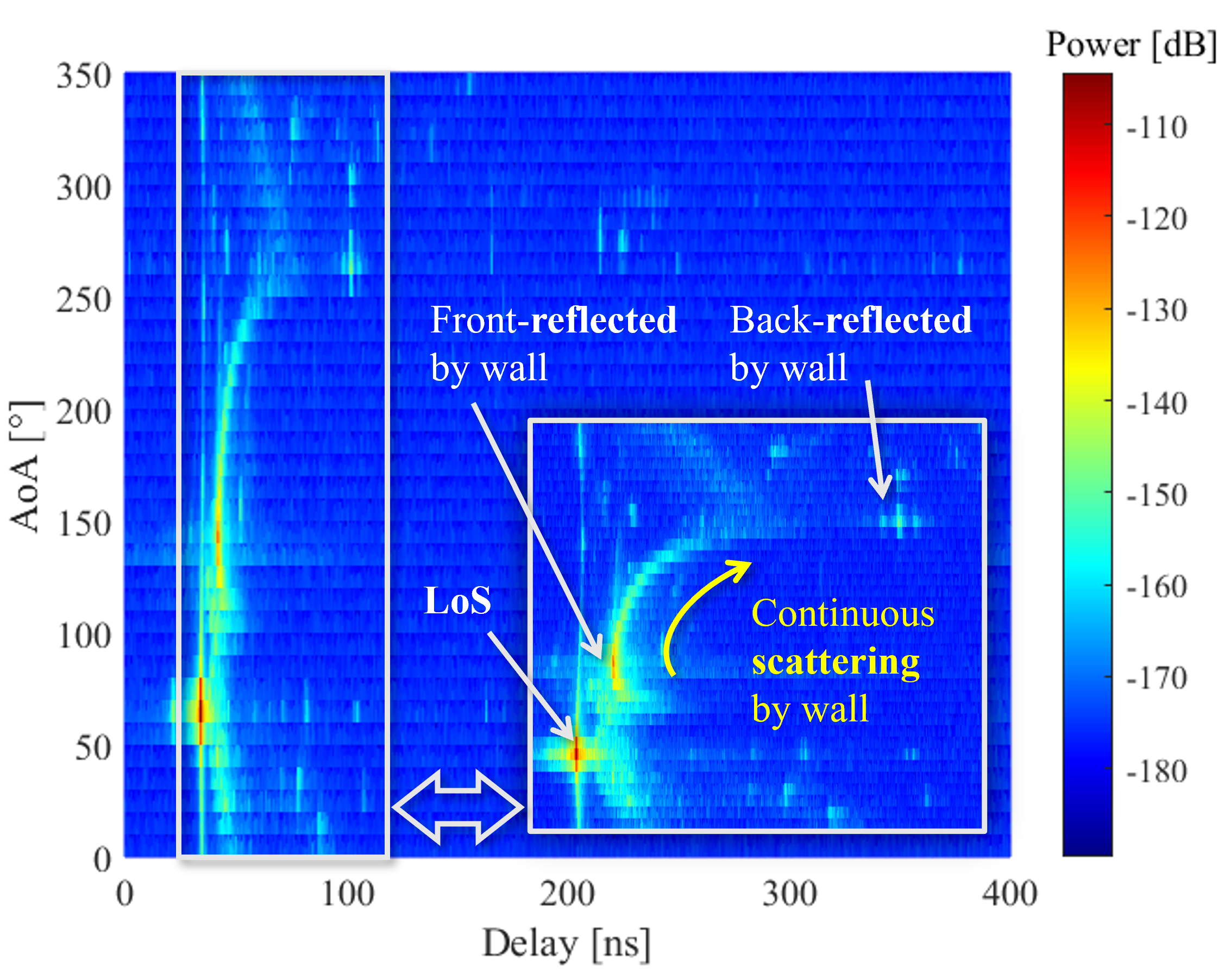}
    \caption{PDAP analysis at Rx1.}
    \label{fig:pdap_pt1}
\end{figure}

\subsection{Measurement Deployment}

The measurement is carried out on a university campus bounded by buildings and streets, which is described in~\cite{wang2023terahertz}. As shown in Fig.~\ref{fig:environment}, the street is parallel to Nanyang Road, and is between Wenbo building and a small parking lot.
The measurement deployment is illustrated in Fig.~\ref{fig:deployment}. We set the west as the horizontal angle $\varphi$ equal to 0$^\circ$, and the value of $\varphi$ increases in the clockwise direction. Tx is located at the northwest corner of the street beside Wenbo Building, and points to the southeast, with the horizontal angle fixed at $\varphi$ = 230$^\circ$. The positions of Rx are linearly distributed along the streets.

\begin{figure}[!t]
    \centering
    \includegraphics[width=\linewidth]{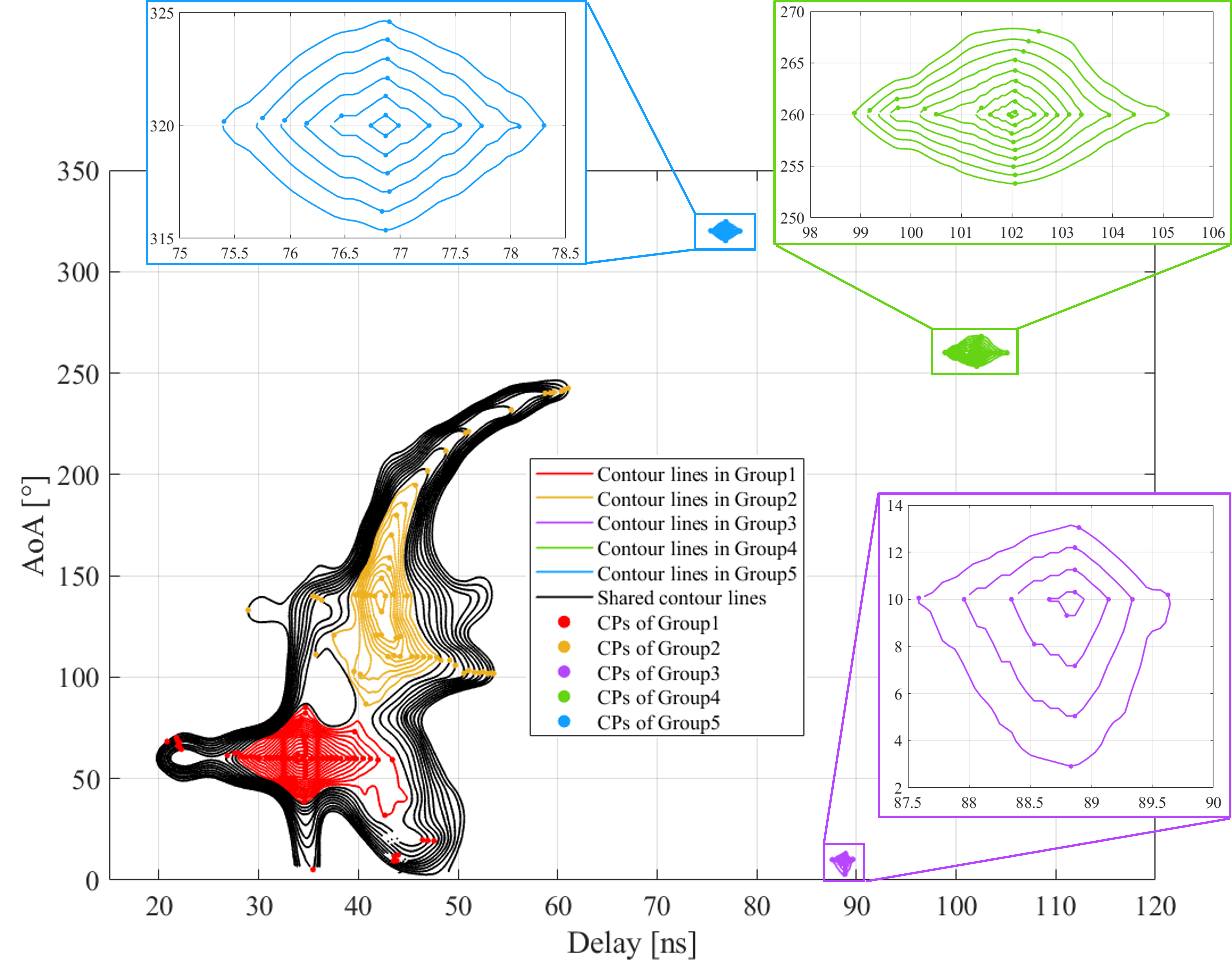}
    \caption{Contour line and CP grouping results of the measurement data based on the proposed algorithm.}
    \label{fig:contour_line_group_summary}
\end{figure}

\begin{figure}[!t]
    \centering
    \begin{subfigure}[Fitted CPs and clustering results.]{
    \includegraphics[width=\linewidth]{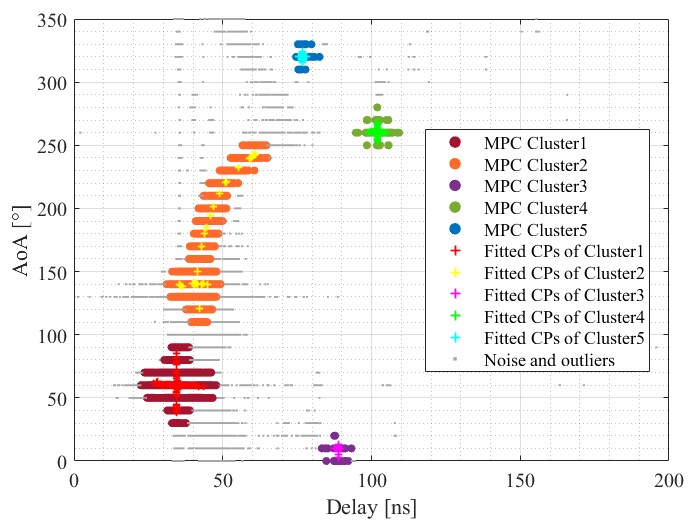}}
    \end{subfigure}
    \\
    \begin{subfigure}[MPC cluster analysis.]{
    \includegraphics[width=0.85\linewidth]{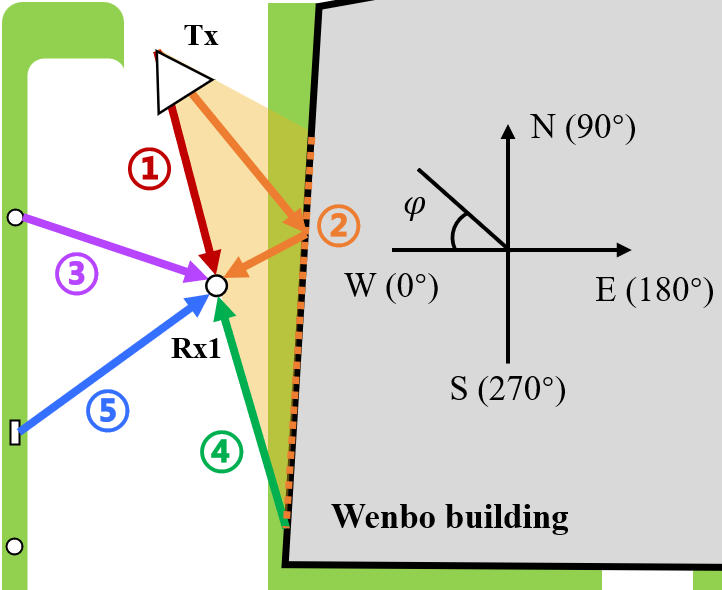}}
    \end{subfigure}
    \caption{Clustering results of the measurement data based on the proposed algorithm.}
    \label{fig:Characteristic_Cluster_Result}
\end{figure}

\subsection{Measurement Result}

\par As illustrated in Fig.~\ref{fig:pdap_pt1} and detailed in~\cite{wang2023terahertz}, three major MPCs are detected, namely, the LoS ray, the front-reflected and back-reflected rays from the wall. Besides, from Rx1 to Rx4, a tail in PDAP is observed between the two wall-reflected rays, with increasing delay, increasing AoA, and decreasing power, which demonstrates the continuous scattering from the wall. This phenomenon occurs since the Tx and Rx positions are close to the wall of Wenbo building so that the power of scattered rays from the wall material is above the noise floor.

\subsection{Scatterer Recognition and Clustering Result} \label{sec:result} 
\par Based on the proposed algorithm elaborated in Section~\ref{sec:algorithm}, the scatterer recognition and clustering result of the measurement data is presented in this part.

\par First, Fig.~\ref{fig:contour_line_group_summary} summarizes the contour lines derived by sweeping the MPC power. The sweeping step affects the number of contour lines and consequently the sampling precision of CPs on the ridge of the cluster. There is no strict requirement for the sweeping step. As long as a moderate value is chosen, e.g. 2\% of the maximum power, the clustering result is not influenced by the sweeping step.
CPs on contour lines are also recognized and assigned into five groups. Second, Fig.~\ref{fig:Characteristic_Cluster_Result}(a) shows the fitted CPs and the final clustering result. The analysis of the physical scatterer corresponding to the clustering result at Rx1 is illustrated in Fig.~\ref{fig:Characteristic_Cluster_Result}(b). To be specific, cluster~1 is dominated by the LoS path, cluster~2 denotes the continuous scattering from the west wall of Wenbo building, cluster~3 is dominated by the reflection from the lamp in the northwest, cluster~4 represents the back-reflection from the wall, and cluster~5 is dominated by the reflection from the traffic sign in the southwest.

\begin{figure}[!t]
    \centering
    \begin{subfigure}[CP fitting result.]{
    \includegraphics[width=\linewidth]{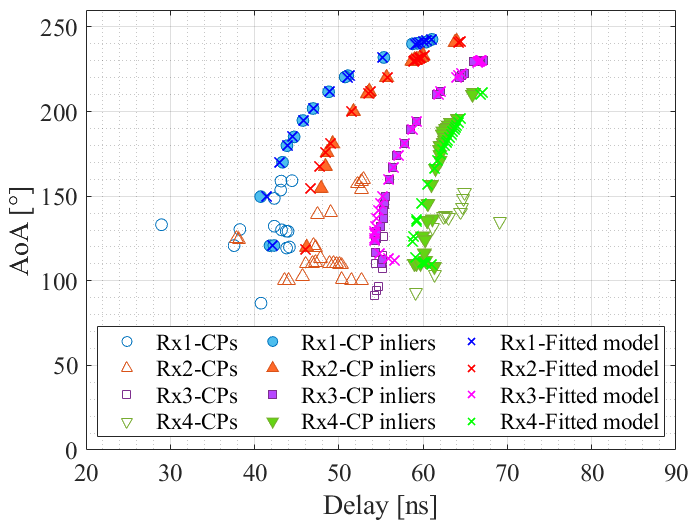}}
    \end{subfigure}
    \\
    \begin{subfigure}[Continuous scatterer reconstruction result.]{
    \includegraphics[width=\linewidth]{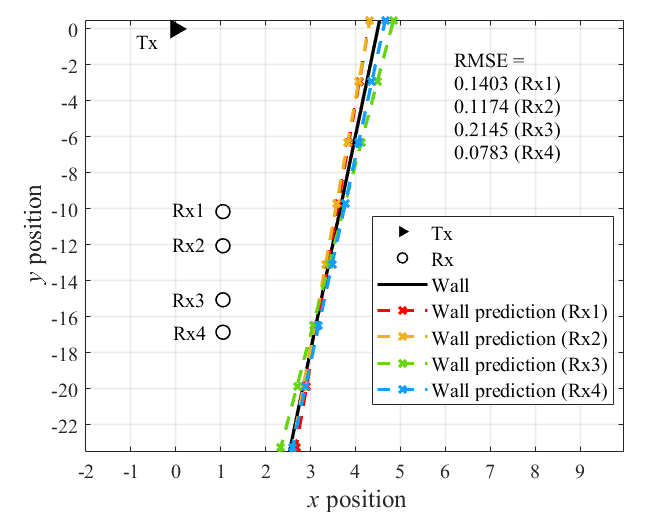}}
    \end{subfigure}
    \caption{Parameter estimation based on CP fitting and continuous scatterer recognition result.}
    \label{fig:wall_prediction_result}
\end{figure}

\begin{table}[!t]
\centering
\caption{Estimated parameters for continuous scatterer reconstruction, derived by CP fitting.}
\label{tab:scatterer_fitting}
\renewcommand\arraystretch{1.2}
\resizebox{\linewidth}{!}{
\begin{tabular}{|c|c|c|c|c|}
\hline
Parameter & $d_{\rm LoS,prior}$ [m] & $d_{\rm LoS}$ [m] & 
$d_{\perp}$ [m] & $\theta$ [$^\circ$] \\ \hline
Rx1 & 10.35 & 10.3329 & 4.3406 & 9.8698 \\ \hline
Rx2 & 12.09 & 12.0412 & 4.3293 & 9.1387 \\ \hline
Rx3 & 14.91 & 14.9281 & 4.8428 & 10.0218 \\ \hline
Rx4 & 16.78 & 16.4544 & 4.6534 & 8.5517 \\ \hline
\end{tabular}
}
\end{table}

\par Specially, we further analyze the reconstruction of the building's wall derived by the fitting result of CPs in cluster~2. The fitting result in Fig.~\ref{fig:wall_prediction_result}(a) derives the estimated parameters for continuous scatterer reconstruction, as summarized in Table~\ref{tab:scatterer_fitting} from Rx1 to Rx4. According to the analytical model in Fig.~\ref{fig:analytical_model}, the geometrical result is illustrated in Fig.~\ref{fig:wall_prediction_result}(b), where the position of the Tx is regarded as the origin. The root mean square error (RMSE) of the wall prediction result is 0.1403~m, 0.1174~m, 0.2145~m, and 0.0783~m for Rx1 to Rx4, respectively. Therefore, the proposed algorithm achieves decimeter-level precision in identifying wide-spread objects, which validates the physical interpretability of the clustering result.

\section{Performance Evaluation and Comparison} \label{sec:compare}

\subsection{Clustering Performance Indices}
\par To evaluate the quality of the clustering result, two kinds of validation metrics are normally used, i.e., external indices and internal indices.
An external index measures the agreement between the result given by the clustering algorithm and the prior-known clustering structure (or the true label).
By contrast, an internal index measures the goodness of a clustering result purely based on inherent features of the data set, without having access to any external information.

\par In the scope of channel measurement, the ``true label'' of MPC clusters is hardly accessible for the following reasons.
First of all, the concept of cluster evolves from the observation of MPCs which arrive in the form of clusters at the receiver side~\cite{saleh1987statistical}. Though the concept of cluster is well accepted and applied in statistical channel modeling, the essence of channel responses is still embedded in MPCs rather than clusters. In this sense, theoretically, the ``true label'' of the cluster structure is not applicable.
On the other hand, the ``true label'' (or reference) of the cluster structure can be generated by existing channel models (or simulators)~\cite{chen2021framework}. However, the credibility of the reference is highly dependent on the agreement between the actual measuring environment and the model. As existing models are generalized to be applied to a certain type of scenarios, specific characteristics of the actual environment are typically weakened or omitted. Therefore, the use of existing general models as the reference is not fair.
Furthermore, the measuring data is susceptible to the measurement resolution, the antenna pattern, and the calibration process~\cite{yin2016performance}.

\par Therefore, in this case, it is more suitable to evaluate the clustering result of the proposed algorithm with internal indices, and compare them with the counterparts of conventional clustering algorithms.
The clustering performance of clustering algorithms on measured MPCs are evaluated by the following indices.

\subsubsection{Silhouette Index}
\par The \textit{Silhouette Index (SI)} measures the closeness of each sample to its own cluster and its separation from other clusters. SI for each sample is defined as
\begin{equation}
    S[n] = \frac{b[n]-a[n]}{\max\{a[n],b[n]\}},
\end{equation}
where $a[n]$ denotes the mean distance between the $n$-th sample and all other samples in the same cluster, and $b[n]$ is the mean distance between the $n$-th sample and all other samples in the next nearest cluster. The distance between two samples, $(\tau_i,\varphi_i)$ and $(\tau_j,\varphi_j)$, is calculated as
\begin{equation} \label{eq:normalize_distance}
    d_{ij} = \sqrt{\frac{(\tau_i - \tau_j)^2}{\sigma_\tau^2} + \frac{(\varphi_i - \varphi_j)^2}{\sigma_\varphi^2}},
\end{equation}
where $\sigma_\tau$ and $\sigma_\varphi$ are standard deviations of samples' delay and AoA, respectively.

\par The SI for a set of samples is given as the mean of SI values for each sample, as
\begin{equation}
    \text{SI} = \frac{1}{N}\sum_{i=1}^{N}S[n],
\end{equation}
where $N$ is the number of inner points given by the clustering result.
The score is bounded between -1 and +1. A high SI score indicates that clusters are dense and well-separated.

\begin{figure}[!t]
    \centering
    \begin{subfigure}[Mean SI.]{
    \includegraphics[width=\linewidth]{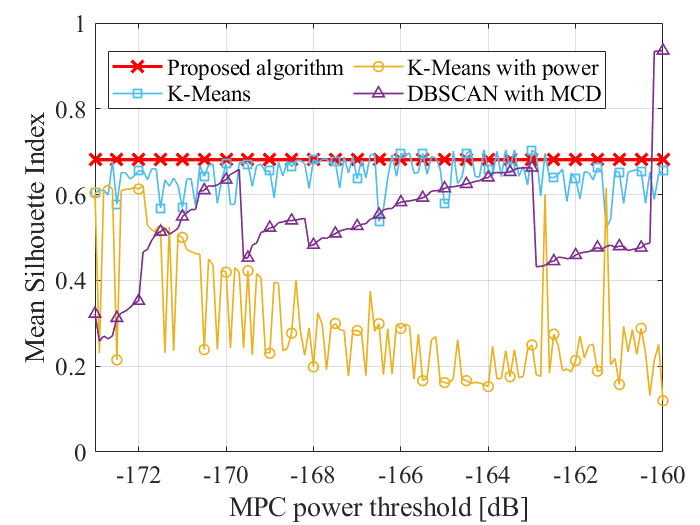}}
    \end{subfigure}
    \\
    \begin{subfigure}[WACC.]{
    \includegraphics[width=\linewidth]{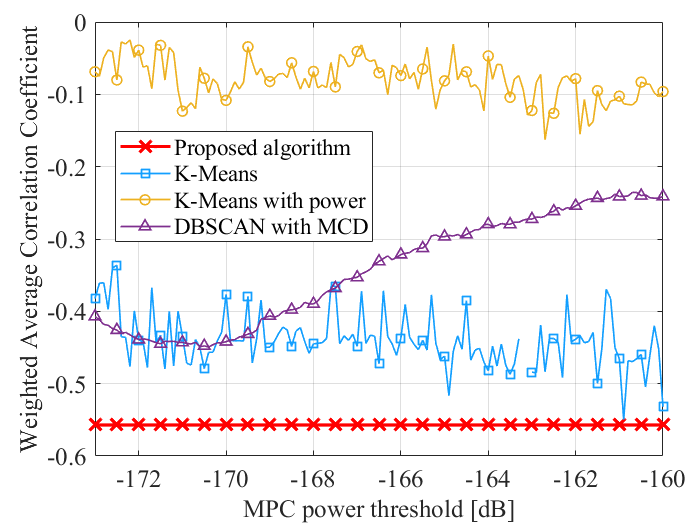}}
    \end{subfigure}
    \caption{Performance comparison of clustering algorithms in (a) mean SI, and (b) WACC. The proposed algorithm is not affected by the power threshold that differentiates valid MPC samples and outliers.}
    \label{fig:performance}
\end{figure}

\subsubsection{Power Gradient Consistency Index}
\par In this part, we propose a new indicator for the performance of MPC clustering, named by the \textit{power gradient consistency index}.
Assuming the most dominant path as the cluster center, the index measures how the power distribution in each cluster adheres to a radially decreasing trend from the center to the periphery, as expected for an MPC cluster where the dominant LoS or reflected path is surrounded by scattering paths with lower power.
\par First, for each cluster, the correlation between the distance and power is estimated by the Spearman correlation coefficient, as
\begin{equation}
    \rho_{k} = \frac{\text{Cov}\left(R(d_k\right),R\left(p_k\right))}{\sigma_{R(d_k)} \cdot \sigma_{R(p_k)}},
\end{equation}
where
\begin{subequations}
\begin{align}
    \text{Cov}\left(R(d_k\right),R\left(p_k\right)) &= \sum_{n=1}^{N_k}(R(d_{k}[n])-\bar{R}_{d_k})(R(p_{k}[n])-\bar{R}_{p_k}), \\
    \sigma_{R(d_k)} &= \sqrt{\sum_{n=1}^{N_k}\left(R(d_{k}[n])-\bar{R}_{d_k}\right)^2}, \\
    \sigma_{R(p_k)} &= \sqrt{\sum_{n=1}^{N_k}\left(R(p_{k}[n])-\bar{R}_{p_k}\right)^2}.
\end{align}
\end{subequations}
$N_k$ is the number of samples in the $k$-th cluster. $d_{k}[n]$ represents the normalized distance between the $n$-th sample and the sample with the highest power in the $k$-th cluster as~\eqref{eq:normalize_distance}. $p_{k}[n]$ denotes the power of the $n$-th sample in the $k$-th cluster. $R(d_{k}[n])$ and $R(p_{k}[n])$ are the rank of $d_{k}[n]$ and $p_{k}[n]$ in $d_{k}$ and $p_{k}$. $\bar{R}_{d_k}$ and $\bar{R}_{p_k}$ are the mean of $R(d_{k}[n])$ and $R(p_{k}[n])$, respectively.
The range of $\rho_k$ is between -1 and +1. When $\rho_k$ is close to +1, $d_k$ and $p_k$ are positively correlated. Otherwise, when $\rho_k$ is close to -1, $d_k$ and $p_k$ are negatively correlated, which is expected for a good clustering result of the PDAP.

\par It appears that when the amount of samples in a cluster $N_k$ is small, the value of $\rho_k$ is prone to extreme values. In this case, though the correlation coefficient $\rho_k$ appears to be significant, it is merely attributed to the small sample size $N_k$ rather than the clustering performance.
Therefore, when we evaluate the power gradient consistency of the overall cluster result with the correlation coefficient of each cluster $\rho_k$, we take into consideration the number of clusters and the number of samples in each cluster.
For fair contribution to the indicator of the overall clustering result, correlation coefficients of each cluster are weighted by its cluster size, so that clusters with more samples have greater influence on the overall indicator. As a result, the power gradient consistency is assessed by the weighted average correlation coefficient (WACC), as
\begin{equation}
    \text{WACC} = \frac{\sum_{k=1}^{N_{c}} N_{k} \cdot \rho_{k}}{_{k=1}^{N_{c}} N_{k}},
\end{equation}
where $N_{k}$ denote the number of samples in the $k$-th cluster, $N_{c}$ is the number of clusters, and $\rho_{k}$ is the correlation coefficient of the $k$-th cluster. WACC close to -1 is expected for a good clustering result of the PDAP, representing that the power and the distance to the cluster center are negatively correlated.

\subsection{Comparison with Conventional Clustering Algorithms}

\par In this part, we compare the clustering performance of the proposed algorithm with conventional algorithms, k-means and DBSCAN. The result is shown in Fig.~\ref{fig:performance} and elaborated as follows. The $x$-axis denotes the power threshold that differentiates valid MPC samples and outliers. The proposed algorithm is not sensitive to the choice of the threshold.

\par
\begin{figure}[!tb]
    \centering
    \begin{subfigure}[K-means.]{
    \includegraphics[width=\linewidth]{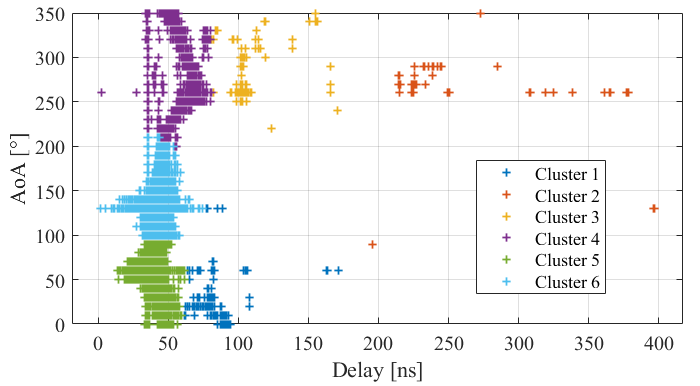}}
    \end{subfigure}
    \\
    \begin{subfigure}[K-means with power.]{
    \includegraphics[width=\linewidth]{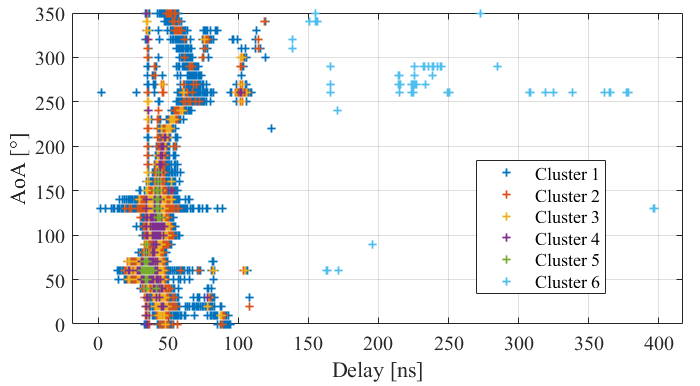}}
    \end{subfigure}
    \\
    \begin{subfigure}[DBSCAN.]{
    \includegraphics[width=\linewidth]{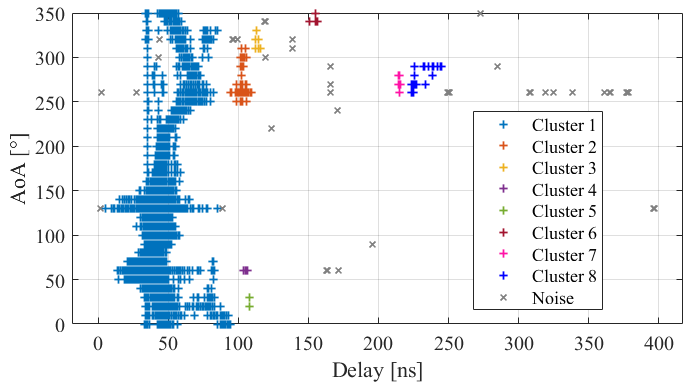}}
    \end{subfigure}
    \caption{Clustering result of the measurement data based on conventional clustering algorithms, with valid MPC power threshold of 170~dB.}
    \label{fig:clustering_conventional}
\end{figure}

 First, k-means is one of the most classic clustering algorithms~\cite{macqueen1967some}. Adaption of it to MPC clustering focuses on the design of the distance metric between two MPC samples. For instance, the weighted distance between $P_{i}(\tau_{i},\varphi_{i})$ and $P_{j}(\tau_{j},\varphi_{j})$ is represented by
\begin{equation}\label{eq:d_km}
    d(P_i,P_j) = \sqrt{w_{\tau}(\tau_{i}'-\tau_{j}')^2+w_{\varphi}(\varphi_{i}'-\varphi_{j}')^2},
\end{equation}
where $\tau_{i}'$ and $\varphi_{i}'$ denote normalized delay and angle, respectively. $w_{\tau}$ and $w_{\varphi}$ represent the weight of temporal and angular differences.
Besides, for fair comparison with other clustering algorithms, we use the elbow method~\cite{bholowalia2014ebk,marutho2018determination} to choose the optimal number of clusters in k-means. 
In Fig.~\ref{fig:performance}, k-means is an effective clustering algorithm, whose performance, in both average SI and WACC, is second only to the proposed algorithm. However, as shown in Fig.~\ref{fig:clustering_conventional}(a), the algorithm is unable to identify outliers, and does not provide any physical interpretation of the clustering result.

\begin{table*}[!tb]
\centering
\caption{Comparison between clustering algorithms.}
\label{tab:compare_algorithm}
\renewcommand\arraystretch{1.1}
\begin{tabular}{|c|c|c|c|c|}
\hline
Properties & \textbf{K-means} & \textbf{K-means with power} & \textbf{DBSCAN} & \textbf{Proposed algorithm} \\ \hline
Attributes & delay, angle & delay, angle, power & delay, angle & delay, angle, power \\ \hline
Cluster shape & spherical & sphrical & arbitrary & arbitrary \\ \hline
Prior knowledge & cluster number & cluster number & self-adaptive & self-adaptive \\ \hline
Outliers sensitivity & sensitive & sensitive & self-adaptive & self-adaptive \\ \hline
\multicolumn{1}{|l|}{Valid power threshold sensitivity} & sensitive & sensitive & sensitive & self-adaptive \\ \hline
Physical meaning & N & N & N & Y \\ \hline
Scatterer identification & N & N & N & Y \\ \hline
\end{tabular}
\end{table*}

\par Obviously the distance metric in~\eqref{eq:d_km} does not fully utilize the attributes in PDAP. This can be improved by including the power values in the distance, for instance,
\begin{equation}
    d(P_i,P_j) = \sqrt{w_{\tau}(\tau_{i}'-\tau_{j}')^2+w_{\varphi}(\varphi_{i}'-\varphi_{j}')^2}+\alpha\vert p_i-p_j\vert,
\end{equation}
where $\alpha$ controls the significance of the difference in power $p_i$ and $p_j$. However, as illustrated in Fig.~\ref{fig:performance} and Fig.~\ref{fig:clustering_conventional}(b), this distance metric does not accord with the intrinsic property of MPC clusters where MPCs in one cluster have descending power values as they deviate from the cluster center. Instead, the algorithm tends to perform radial slicing on the real MPC cluster.

\par Except for k-means, DBSCAN is frequently used for MPC clustering based on PDAPs derived from channel measurement. Besides, the distance metric is tailored to MPC samples by the MCD in~\cite{chen2021channel}. Compared with k-means, the algorithm can identify outliers, and is more flexible in the shape of the cluster. However, as shown in Fig.~\ref{fig:clustering_conventional}(c), it cannot handle the situation when MPC clusters become conterminous due to the spread. As a result, the performance of DBSCAN is significantly sensitive to the valid MPC power threshold.

\par In Table~\ref{tab:compare_algorithm}, we summarize properties of k-means, k-means with power, DBSCAN, and the proposed algorithm. In contrast to conventional clustering algorithms, the proposed algorithm effectively exploits power values in the PDAP. In particular, by characterizing the intrinsic of MPC clusters, which is analogous to mountain peaks in physical geography, the proposed algorithm provides significant physical evidence for the clustering result. Besides, by utilizing contour lines, the tree structure and the definition of characteristic points, the algorithm deconstructs the PDAP and therefore is self-adaptive in adjusting parameters and eliminating noise and outliers. Furthermore, the clustering result also provides clues for identifying physical scatterers that contribute to the cluster.

\section{Conclusion} \label{sec:conclusion}

In this paper, we have developed a novel metaphor that interprets features of MPC attributes in the PDAP as topographic concepts. In light of the interpretation, we have proposed a novel MPC clustering algorithm that disassemble the PDAP by constructing and organizing contour lines in a hierarchical structure with full utilization of MPC attributes. CPs that indicate the skeleton of MPC clusters, are identified as convex points on contour lines, which direct the clustering of MPCs and are fitted by analytical models that associate MPCs with physical scatterer locations. Based on the channel measurement result in an outdoor street, the algorithm is able to identify both single-point and wide-spread scatterers without prior knowledge, with the average RMSE of 0.1~m. Besides, a new clustering performance index, named as the power gradient consistency index, is proposed to assesses whether the clustering result captures the intrinsic property of MPC clusters that the dominant high-power path is surrounded by lower-power paths. In terms of the new index and the conventional Silhouette index, the proposed algorithm has better clustering performance compared with traditional clustering algorithms. Furthermore, the algorithm is insensitive to the valid MPC power threshold and does not require user-specific parameters.


\bibliographystyle{IEEEtran}
\bibliography{bibliography}

\end{document}